%% file: ms.tex
\documentclass[usenatbib,useAMS,a4paper]{mn2e}
\usepackage{txfonts}
\DeclareSymbolFont{cmletters}{OML}{cmm}{m}{it}
\DeclareMathSymbol{v}{\mathalpha}{cmletters}{"76}
\usepackage{graphicx}

\newcommand{\useiop}{-3}

\newcommand{\shortauthors}[1]{}
\newcommand{\shorttitle}[1]{}
\newcommand{\altaffiltext}[2]{}
\newcommand{\eqref}[1]{(\ref{#1})}
\input{macros}
\usepackage{ifthen}
\usepackage[usenames,dvipsnames]{color}

\definecolor{MyDarkBlue}{rgb}{0,0.08,0.7}

\defcitealias{tch08}{TMN08}
\defcitealias{tch09}{TMN09}
\defcitealias{kom09}{K09}
\defcitealias{bz77}{BZ77}

\newcommand{\OmegaH}{\Omega_\mathrm{H}}

\newcommand{\rH}{r_\mathrm{H}}

\newcommand{\const}{{\rm constant}}

\newcommand{\gdet}{\sqrt{-g}}

\newcommand{\avg}[1]{\ensuremath{\langle#1\rangle}} 

\newcommand{\MdotH}{\dot M}

\newcommand{\etabh}{\eta_{\rm BZ}}

\newcommand{\phibh}{\phi_{\rm BH}}

\newcommand{\cut}[1]{\hbox{}}

\shortauthors{}
\shorttitle{}

\ifthenelse{\equal{\useiop}{-3}}{ 

\author[A.~Tchekhovskoy,
R.~Narayan and
J.~C.~McKinney]
{Alexander Tchekhovskoy$^1$\thanks{\hbox{E-mail: atchekho@princeton.edu~(AT)}},
Ramesh Narayan$^2$,
Jonathan C. McKinney$^3$
\\
  $^1$Center for Theoretical Science, Jadwin Hall, Princeton University, Princeton,
  NJ 08544; Princeton Center for Theoretical Science Fellow \\
  $^2$Institute for  Theory and Computation, Harvard-Smithsonian Center for Astrophysics,
 60 Garden Street, MS 51, Cambridge, MA 02138, USA
\\
 $^3$Kavli Institute for Particle Astrophysics and Cosmology, Stanford University, P.O. Box 20450, MS 29,
Stanford, CA 94309; Einstein Fellow }
}{
}
\begin{document}
\label{firstpage}

\title[Jets from Magnetically Arrested BH Accretion]%
{Efficient Generation of Jets from Magnetically
  Arrested Accretion on a Rapidly Spinning Black Hole}

\ifthenelse{\equal{\useiop}{-3}}{ 
\date{Accepted . Received ; in original form }
\pagerange{\pageref{firstpage}--\pageref{lastpage}} \pubyear{2011}
\maketitle
}

\begin{abstract}

  We describe global, 3D, time-dependent, non-radiative,
  general-relativistic, magnetohydrodynamic simulations of
  accreting black holes (BHs).  {The simulations are designed to
  transport a large amount of magnetic flux to the center, more than
  the accreting gas can force into the BH}.  The excess magnetic flux remains outside the
  BH, impedes accretion, and leads to a magnetically arrested disc.
  We find powerful outflows.  For a BH with spin parameter $a=0.5$,
  the efficiency with which the accretion system generates outflowing
  energy in jets and winds is $\eta\approx 30$\%.  For $a=0.99$, we
  find $\eta\approx 140$\%, which means that more energy flows out of
  the BH than flows in. The only way this can happen is by extracting 
  spin energy from the BH.  Thus the $a=0.99$ simulation represents an unambiguous
  demonstration, within an astrophysically plausible scenario, of the
  extraction of net energy from a spinning BH via the
  Penrose-Blandford-Znajek mechanism.  We suggest that magnetically
  arrested accretion might explain observations of active galactic nuclei with apparent
  $\eta\approx {\rm few}\times100\%$.

\end{abstract}

\ifthenelse{\equal{\useiop}{-2}}{
\end{frontmatter}
}{}

\ifthenelse{\equal{\useiop}{-3}}{ 
}{}
\ifthenelse{\equal{\useiop}{3}}{
  }{}
  
\ifthenelse{\equal{\useiop}{0}}{
{
  }
}


\section{Introduction}
\label{sec:introduction}

\begin{figure*}
  \begin{center}
    \includegraphics[width=0.88\textwidth]{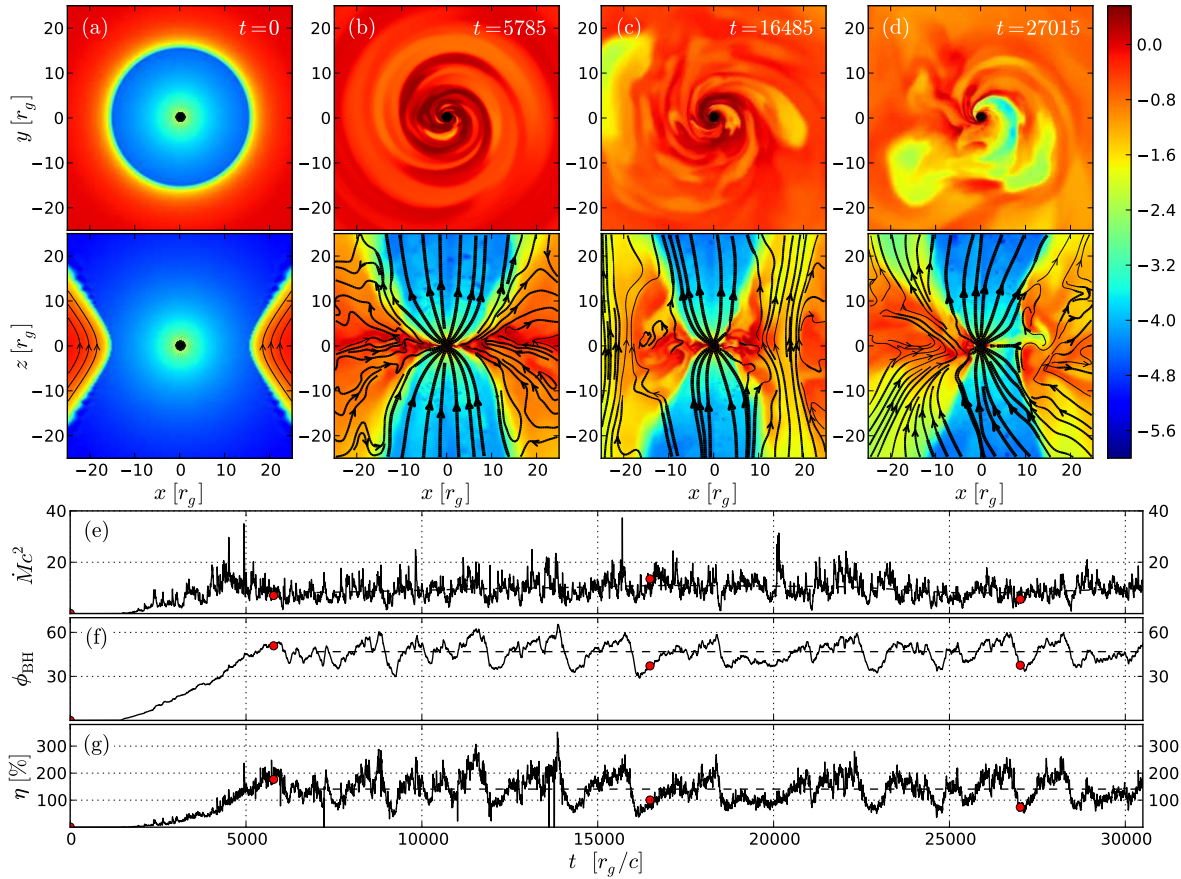}
  \end{center}
  \caption{Shows results from the fiducial GRMHD simulation A0.99fc
    for a BH with spin parameter $a=0.99$. The accreting gas in this
    simulation settles down to a magnetically arrested state of
    accretion.  [Panels (a)-(d)]: The top and bottom rows show,
    respectively, equatorial ($z=0$) and meridional ($y=0$) snapshots
    of the flow, at the indicated times. Colour represents the
    logarithm of the fluid frame rest-mass density, $\log_{10}\rho
    c^2$ (red shows high and blue low values; see colour bar), filled
    black circle shows BH horizon, and
    black lines show field lines in the image plane.  [Panel
    (e)]: Time evolution of the rest-mass accretion rate, $\MdotH
    c^2$. 
    The
    fluctuations are due to turbulent accretion and are
    normal. The long-term trends, which we show with a Gaussian smoothed (with width
    $\tau=1500r_g/c$) accretion rate, $\avg{\MdotH}_\tau
    c^2$, are small (black dashed line).
    [Panel (f)]: Time evolution of the
    large-scale magnetic
    flux, $\phi_{\rm BH}$, threading the BH horizon,
    normalized by
    $\avg{\MdotH}_{\tau}$ .  The magnetic flux
    continues to grow until $t\approx 6000 r_g/c$. Beyond this time,
    the flux saturates and the accretion is magnetically arrested.
    (Panels (c) and (d) are during this period). The large amplitude
    fluctuations are caused by quasi-periodic accumulation and escape
    of field line bundles in the vicinity of the BH. [Panel (g)]
    Time evolution of the energy outflow efficiency  $\eta$
    (defined in eq.~\ref{eq:eta2} and here
    normalized to $\avg{\MdotH}_\tau c^2$).  Note the large fluctuations in
    $\eta$, which are well-correlated with corresponding fluctuations
    in $\phi_{\rm BH}$.  Dashed lines in panels (f) and (g)
    indicate time averaged
    values, $\avg{\phibh^2}^{1/2}$ and $\avg\eta$,
    respectively.  The average $\eta$ is clearly greater than $100\%$,
    indicating that there is a net energy flow out of the BH.
  }
\label{fig:qtyvst}
\end{figure*}

Relativistic jets are a common feature of accreting black holes (BHs).
They are found in both stellar-mass BHs \citep{rem06} and
supermassive BHs in active galactic nuclei (AGN, \citealt{t02}). Jets
can be very powerful, with their energy output sometimes exceeding
the Eddington limit of the BH. This suggests
an efficient mechanism for their production.

In seminal work, \citet{Penrose1969} showed that a spinning BH has free
energy that is, in principle, available to be tapped.  This has led to
the popular idea that the energy source behind relativistic jets is
the rotational energy of the accreting BH. \citet[hereafter \citetalias{bz77}]{bz77} came up with
an astrophysical scenario in which this could be achieved. In their
picture, magnetic field lines are kept confined around the BH by an
accretion disc.  The rotation of space-time near the BH twists
these lines into helical magnetic springs which expand under their own
pressure and accelerate any attached plasma. In the
process, energy is extracted from the spinning BH and is transported
out along the magnetic field, making a relativistic jet.  The BZ
mechanism is a promising
idea since magnetic fields are common in astrophysical accretion
discs and so the requirements for this mechanism are easily met.  

In the BZ mechanism, the rate at which rotational energy of the BH is
extracted -- the BZ power $P_{\rm BZ}$ -- 
is given in Gaussian-cgs units by (\citetalias{bz77}, \citealt*{tch10a})
\begin{equation}
P_{\rm BZ} = \frac{\kappa}{4\pi c} \; \OmegaH^2 \, \Phi_{\rm BH}^2 \,
f(\OmegaH), 
\label{eq:cgspower}
\end{equation}
where $\kappa$ is a numerical constant whose value depends on the
magnetic field geometry (it is 0.053 for a split monopole geometry and
0.044 for a parabolic geometry), $\OmegaH=ac/2\rH$ is the angular
frequency of the BH horizon, $\Phi_{\rm BH}=(1/2)\int_\theta\int_\varphi |B^r|
dA_{\theta\varphi}$ is the magnetic flux threading one hemisphere of
the BH horizon (the integral is over all $\theta$, $\varphi$ at the BH horizon, and the
factor of $1/2$ converts it to one hemisphere),
$dA_{\theta\varphi}=\gdet d\theta d\varphi$ is an area element in the
$\theta{-}\varphi$ plane, and $g$ is the determinant of the metric.
Here $a$ is the dimensionless BH spin parameter (sometimes also called
$a_*$), $\rH=r_g(1+\sqrt{1-a^2})$ is the radius of the horizon,
$r_g=GM/c^2$ is the gravitational radius of the BH, and $M$ is the BH
mass.
A simpler version of equation (\ref{eq:cgspower}) with
$P\propto a^2$ was originally derived by \citetalias{bz77} in the limit
$a\ll 1$.  \citet{tch10a} showed that the modified form written here,
with $f(\OmegaH)=1$, is
accurate even for large spins up to $a \approx 0.95$, 
while for yet larger spins, they gave a more accurate 6th order
approximation, $f(\OmegaH) \approx 1+1.38(\OmegaH
r_g/c)^2-9.2(\OmegaH r_g/c)^4$.

Using equation~\eqref{eq:cgspower}, let us define the efficiency with which 
the BH generates jet power, $\eta_{\rm  BZ}$, as the ratio of the
time-average electromagnetic power that flows out of the BH,
$\avg{P_{\rm BZ}}$, to the time-average rate at which rest-mass energy flows into
the BH, $\avg{\dot M} c^2$, 
\begin{equation}
  \label{eq:eta}
  \etabh \equiv \frac{\avg{P_{\rm BZ}}}{\avg{\MdotH} c^2}\times100\% = 
  \frac{\kappa}{4\pi c} \; \left(\frac{\OmegaH r_g}{c}\right)^2
  \avg{\phibh^2} \, f(\OmegaH)\times 100\%,
\end{equation}
where $\phibh = \Phi_{\rm BH}/\left(\avg{\MdotH}r_g^2 c\right)^{1/2}$
is the dimensionless magnetic flux threading the BH and $\avg{...}$
is a time-average.
Thus the efficiency with which a
spinning BH can generate jet power depends on BH spin $a$ via
the angular frequency $\OmegaH$ and on the dimensionless magnetic flux $\phibh$.
The strength of $\phibh$ is very uncertain.

It is generally agreed that $\phibh$ is non-zero, since magnetic flux
is transported to the accreting BH by turbulent accretion. However,
the key elements of this process are not agreed upon
\citep{lubow1994,spruit_uzdensky_2005,rl08,bhk09,cao11}.
This leads to a large uncertainty in the value of $\etabh$.
Recent time-dependent general relativistic magnetohydrodynamic (GRMHD)
numerical simulations have found a rather low efficiency,
$\etabh\lesssim20\%$,
even when the central BH is nearly maximally spinning
\citep{mck05,dev05a,hk06,bb11}.  With such a modest efficiency it is not
clear that we are seeing energy extraction from the BH. The jet
power could easily come from the accretion disc \citep[see][]{ga97,lop99}.  

Observationally, there are indications that some AGN in the universe
may have extremely {efficient} jets that require $\eta\gtrsim100\%$
\citep{rs91,ghi_blazars_2010,fernandes_agnjetefficiency_2010,mcnamara_agnjetefficiency_2010,punsly2011}.  A non-spinning BH usually has $\eta < 10\%$,
and might under special circumstances have $\eta \approx$ tens of percent
(e.g., \citealt{nia03}). However,
a non-spinning BH can never give $\eta > 100\%$, since this
requires the system to produce more energy
than the entire rest mass energy supplied by accretion. Values
of $\eta>100\%$ are
possible only by extracting energy from the spin of the BH. 
Thus, taken at face value, any robust observation of $\eta>100\%$
in an AGN implies that the Penrose/BZ
process must be operating.  This raises
the following important question: Is it possible to show via a
numerical simulation, an astrophysically plausible BH accretion
scenario that gives jet efficiency $\eta>100\%$? To our knowledge, this
has not been demonstrated with a GRMHD simulation.

\begin{table*}
\begin{center}
\caption{Simulation details}
\begin{minipage}{\textwidth}
\begin{center}
\begin{tabular}{lccc@{$\quad$}c@{$\quad$}c@{$\quad$}c@{$\quad$}c@{$\quad$}c@{$\quad$}ccc}
\hline
Name$^a$       &  $a$  & $\eta\ [\%]$ &$\Delta\varphi$&   Resolution
($N_r\times N_\theta\times N_\varphi$) &$\xi_\mathrm{max}$ & $r_{\rm in}/r_g$ & $r_{\rm max}/r_g$ &
$h/r$ at $r_{\rm max}$ & $r_{\rm br}/r_g$ & $t_{\rm run}\ [r_g/c]$ & $t_{\rm avg}\ [r_g/c]$\\
\hline
A0.5  &0.5 & $30\pm5$ &   $\pi$   & $288\times128\times32$ & $25$ & $15$ & $34.475$& $0.2$ & $200$ & ($0$; $13,\!095$) & ($10,\!300$; $13,\!095$) \\ 
A0.99 &$0.99$& $145\pm15$  &   $\pi$   & $288\times128\times32$ & $25$ & $15$ & $34$ &$0.2$ & $200$ & ($0$; $13,\!370$) & ($6,\!000$; $13,\!370$) \\ 
A0.99f     &$0.99$&$150\pm10$&  $2\pi$   & $288\times128\times64$ &$250$ & $15$ & $34$ &$0.2$ & $1000$ & ($0$; $14,\!674$) & ($7,\!000$; $14,\!674$) \\ 
A0.99fh&$0.99$&$135\pm10$ &  $2\pi$ & $288\times128\times128$&$250$ & $15$ & $34$ & $0.2$ & $1000$ & ($14,\!674$; $30,\!500$) & ($14,\!674$; $30,\!500$) \\ 
A0.99fc  &$0.99$&$140\pm15$ & $2\pi$& $288\times128\times\{64,128\}$
& $250$& $15$ & $34$&$0.2$   &$1000$  & ($0$; $30,\!500$) & ($7,\!000$; $30,\!500$) \\
\hline
\label{tab1}
\end{tabular}
\end{center}
\vspace{-0.5cm}
$^a$ Model A0.99fh is similar to A0.99f but with $N_\varphi$
increased by a factor of two at $t=14,\!674r_g/c$.  Model A0.99fc is comprised of models A0.99f and A0.99fh.
\end{minipage}
\end{center}
\end{table*}

{Here we describe numerical simulations in which we arrange
our setup such that the accreting BH receives as much large-scale
magnetic flux as 
can be pushed into the BH by accretion. In nonradiative MHD, the limiting flux is
proportional to $\MdotH$. In fact, we supply more flux than this, 
so some of the flux remains outside the BH
where it impedes the accreting gas, leading to a ``magnetically
arrested disc'' (MAD, \citealt{nia03}, see also
\citealt{bkr74,bkr76,igu03}).}   The goal of the present simulations
is to maximize $\phibh$ 
and to make the jet efficiency as large as possible.
As we show below, we do obtain
larger efficiencies than reported in previous numerical experiments.
Most interestingly, we find efficiencies greater than $100\%$ for a rapidly spinning
BH.  
These experiments
are the first demonstration of net energy extraction from spinning BHs
via the Penrose/BZ mechanism 
in an astrophysically plausible setting.  In
\S\ref{sec:problem-setup}, we discuss our numerical method, the
physics of MAD accretion and our problem setup, and
in \S\ref{sec:results} we discuss the results and conclude.

\section{Numerical Method and Problem Setup}
\label{sec:problem-setup}

We have carried out time-dependent simulations of BH
accretion for two values of BH spin (see Table~\ref{tab1}). 
We use the GRMHD code {\sc HARM} \citep{gam03,mck04}
with recent improvements \citep{mck06jf,tch07,tch09,mb09}. The
numerical method conserves mass, angular momentum and energy 
to machine precision. We neglect radiative losses, so the simulations
correspond to an ADAF mode of accretion \citep{nm08}.

The simulations are carried out in spherical polar coordinates  
modified to concentrate resolution in the collimating polar jet
and in the equatorial disc.
We use Kerr-Schild horizon-penetrating coordinates. We place the inner
radial boundary inside the BH outer horizon (but outside the inner
horizon), which ensures that no signals can propagate through the horizon
to the BH exterior. The outer radial boundary is at $r=10^5r_g$, which
exceeds the light travel distance for the duration of the simulation.
Thus both boundaries are causally disconnected.  We use a
logarithmically spaced radial grid, $dr/(r-r_0)=\const$ for $r\lesssim
r_{\rm br}$ (see Table~\ref{tab1} for values of $r_{\rm br}$), 
where we choose $r_0$ so that
there are $9$ grid cells between the inner radial boundary and the BH
horizon. For $r\gtrsim r_{\rm br}$, the radial
grid becomes progressively sparser, $dr/r=4(\log r)^{3/4}$, with a
smooth transition at $r_{\rm br}$.  At the poles, we use the standard
reflecting boundary conditions, while in the azimuthal direction, we
use periodic boundary conditions.  In order to prevent the
$\varphi-$extent of cells near the poles from limiting the time step,
we smoothly deform the grid a few cells away from the pole so
as to make it almost cylindrical near the BH horizon; this speeds up the
simulations by a factor $\gtrsim5$.

Numerical MHD schemes cannot handle vacuum. Therefore, whenever 
the fluid-frame rest-mass energy density, $\rho c^2$, falls
below a density floor $\rho_{\rm floor} c^2=p_{\rm mag}/\xi_{\rm max}$, where $p_{\rm mag}$ is the magnetic pressure in the fluid
frame, or when the internal
energy density, $u_{\rm g}$, falls below $u_{\rm g,floor}=0.1\rho_{\rm
floor}c^2$, we add mass or internal energy in the frame of a local zero
angular momentum observer so as to make $\rho = \rho_{\rm
floor}$ or $u_{\rm g} = u_{\rm g, floor}$ \citep{mb09}. The factor
$\xi_{\rm max}$ sets the maximum possible Lorentz factor of the jet
outflow.  To investigate the effect of this factor on jet efficiency, we
have tried two values, $\xi_{\rm max} = 25$, $250$. There is little difference in
the results.  In any case, we track the amount of mass and internal
energy added in each cell during the course of the simulation and we 
eliminate this contribution when calculating
mass and energy fluxes.

Model A0.99f (Table~\ref{tab1}) uses a resolution of
$288\times128\times64$ along $r$-, $\theta$-, and $\varphi$-,
respectively, and a full azimuthal wedge, $\Delta\varphi = 2\pi$. This
setup results in a cell aspect ratio in the equatorial region, $\delta
r:r\delta\theta:r\delta\varphi\approx1:0.4:5$. To check convergence
with numerical resolution, at $t=14,\!674r_g/c$, well after the model
reached steady-state, we dynamically increased the
number of cells in the azimuthal direction by a factor of $2$.  We
refer to this higher-resolution simulation as model A0.99fh and to
A0.99f and A0.99fh combined as model A0.99fc.
We also ran model A0.99 with
a smaller azimuthal wedge, $\Delta\varphi=\pi$. 
We find that the time-averaged jet efficiencies of the four A0.99xx models
agree to within statistical
measurement uncertainty (Table~\ref{tab1}), indicating that our results are
converged with respect to azimuthal resolution and wedge size.

Our fiducial model A0.99fc starts with a rapidly spinning BH 
($a=0.99$) at the center of an equilibrium hydrodynamic torus
\citep{chakrabarti1985,dev03a}. The inner edge of the torus is at $r_{\rm
  in}=15r_g$ and the pressure maximum is at $r_{\rm max}=34r_g$ (see
Fig.~\ref{fig:qtyvst}a).  At $r=r_{\rm max}$ the initial torus has
an aspect ratio $h/r\approx0.2$ and fluid frame density $\rho=1$ (in
arbitrary units). The torus is seeded with a weak large-scale
poloidal magnetic field (plasma $\beta\equiv p_{\rm gas}/p_{\rm
  mag}\ge100$). This configuration is unstable to the magnetorotational
instability \citep[MRI,][]{bal91} which drives MHD turbulence and
causes gas to accrete.
The torus serves as a reservoir of mass and magnetic field for the
accretion flow.

Equation (\ref{eq:cgspower}) shows that the BZ power is directly
proportional to the square of the magnetic flux at the BH horizon,
which is determined by the large-scale poloidal magnetic flux supplied
to the BH by the accretion flow.  The latter depends on the initial
field configuration in the torus. {Usually, the initial field is chosen
to follow isodensity contours of the torus, e.g., the magnetic flux
function is taken as $\Phi_1(r,\theta)=C_1\rho^2(r,\theta)$, where
  the constant factor $C_1$ is tuned to achieve the desired
minimum value of $\beta$
in the torus, e.g., $\min\beta=100$.  The resulting
poloidal magnetic field loop is centered at
$r=r_{\rm max}$ and contains a relatively small amount of magnetic flux. If
we wish to have an {efficient} jet, we need a torus with more magnetic
flux, so that some of the flux remains outside
the BH and leads to a MAD state of accretion
\citep{igu03,nia03}.  We achieve this in several steps. We consider a
magnetic flux function, $\Phi(r,\theta) =
r^{5}\rho^2(r,\theta)$, and normalize the magnitude of the
magnetic field at each point independently such that we have $\beta = \const$
everywhere in the torus.  
Using this field, we take the initial magnetic flux
function as $\Phi_2(r,\theta)
=C_2\int_{\theta'=0}^{\theta'=\theta}\int_{\varphi'=0}^{\varphi'=2\pi}
B^{r} dA_{\theta'\varphi'}$ and
tune $C_2$ such that $\min \beta=100$.
This gives a poloidal field loop centered at $r \simeq 300 r_g$.
The loop has a much larger spatial size and more magnetic flux
than the usual initial field loop configuration considered in other
studies. We maintain a nearly uniform radial distribution of $\beta$ for
$r\lesssim300r_g$ which lets us resolve the fastest growing MRI wavelength
with more than $10$ cells over a wide range of radii.}

Note that the above technique of starting with a large amount of
poloidal magnetic flux in the torus is just a convenient trick to
achieve a MAD state of accretion within the short
time available in a numerical simulation.   
{In our simulations magnetic flux is rapidly advected to the center
from a distance $r\lesssim10^2r_g$, whereas
in nature we expect magnetic flux to be advected from a
distant external medium at $r>10^5r_g$ and to grow on the corresponding accretion
time.  The latter time is much
too long to be currently simulated on a computer, hence the need to speed up the
process in the simulations.  Note that field advection
is likely to be more efficient in a thick, ADAF-like disc
rather than in a thin disc
\citep[e.g.,][]{lubow1994,spruit_uzdensky_2005,rl08,cao11}, hence we consider
the present simulations to be a reasonable proxy for real ADAFs in nature.
In any case, the key point is that, regardless of
how a given system achieves a MAD state of accretion
--- whether it is by slow advection of field from
large distances in a real system or
through rapid advection of magnetic flux from short distances in our simulations ---
once the system has reached this state we expect its properties to be
largely insensitive to its prior history.   We thus believe the results
obtained here are relevant to astrophysical objects with MADs.}

\section{Results}
\label{sec:results}

Figure~\ref{fig:qtyvst} shows results from the fiducial simulation
A0.99fc for a rapidly spinning BH with $a=0.99$.  The top two panels
(a) on the left show the initial torus, with purely poloidal magnetic
field. The succeeding panels show how the accretion flow evolves. With
increasing time, the MRI leads to MHD turbulence in the torus which
causes the magnetized gas to accrete on the BH. In the process,
magnetic flux is brought to the center and accumulates around the BH
in an ordered bipolar configuration.  These
field lines are twisted into a helical shape as a result of space-time
dragging by the spinning BH and they carry away energy along twin
jets.

The rate of accretion of rest mass, $\dot{M}(r)$, and rest mass energy,
$F_M(r)\equiv \dot{M}(r)c^2$, at radius $r$ are given by
\begin{equation}
\dot{M}(r) = -\int_\theta \int_\varphi \rho u^r dA_{\theta\varphi} \equiv F_M(r)/c^2 ,
\end{equation}
where $u^r$ is the radial contravariant component of the $4$-velocity, and 
the integral is over all $\theta$, $\varphi$ at fixed $r$. The
negative sign means that the flux is defined to be positive
when rest mass flows into the BH. 
Figure~\ref{fig:qtyvst}(e) shows the rest mass energy flux into the
BH, $F_M(\rH) = \dot{M}(\rH)c^2$, as a function of time (the
flux has been corrected for density floors, see
\S\ref{sec:problem-setup}). Until a time $t\sim2000r_g/c$, the MRI is
slowly building up inside the torus and there is no significant
accretion. After this time, $F_M(\rH)$ steadily grows until it
saturates at $t\sim4000r_g/c$. Beyond this time, the accretion rate
remains more or less steady at approximately $10$ code units
until the end of the simulation at $t\sim 30000r_g/c$.  The
fluctuations seen in $F_M$ are characteristic of turbulent accretion via
the MRI. 

Figure~\ref{fig:qtyvst}(f) shows the time evolution of the
dimensionless magnetic flux $\phibh$ at the BH horizon. Since the
accreting gas continuously brings in new flux, $\phibh$ continues to
grow even after $F_M$ saturates. {However, there is a limit to how much
flux the accretion disc can push into the BH.}  Hence, at $t\sim6000r_g/c$, the flux on the
BH saturates and after that remains roughly constant at a value
around $47$. The corresponding dimensionless
magnetization parameter $\Upsilon$ (\citealt{gam99}; see
\citealt{penna10} for definition) is
$\approx9.5$ (much greater than $1$), indicating that
the flow near the BH is highly magnetized. Panel (b) shows that magnetic fields near
the BH are so strong that they compress the inner accretion disc
vertically and  decrease its thickness. 
The accreting gas, of course, continues to
bring even more flux, but this additional flux remains outside the
BH. Panels (c) and (d) 
show what happens to
the excess flux. Even as the gas drags the magnetic field in, field
bundles erupt outward \citep{igu08}, leaving the
time-average flux on the BH constant. 
For instance, two flux bundles are
seen at $x\sim\pm20r_g$ in Figure~\ref{fig:qtyvst}(c) which originate in
earlier eruption events. Other bundles are similarly 
seen in Figure~\ref{fig:qtyvst}(d).  During
each eruption, the mass accretion rate is partially suppressed,
causing a dip in $\dot{M}c^2$ (Fig.~\ref{fig:qtyvst}e); there is also a
corresponding temporary dip in $\phibh$ (Fig.~\ref{fig:qtyvst}f).
Note that, unlike in 2D (axisymmetric) simulations
\citep[e.g.,][]{pb03}, there is never
a complete halt to the accretion \citep{igu03} and even during flux
eruptions, accretion proceeds via spiral-like structures, as seen in
Figure~\ref{fig:qtyvst}(d).

In analogy with $F_M$, let us define the rate of inward flow of total energy
(as measured at infinity) as follows,
\begin{equation}
F_E(r) = \int_\theta \int_\varphi {T^r}_{\!t}\, dA_{\theta\varphi},
\end{equation}
where ${T^\mu}_{\!\nu}$ is the stress-energy tensor.
\begin{figure}
  \begin{center}
    \includegraphics[width=\columnwidth]{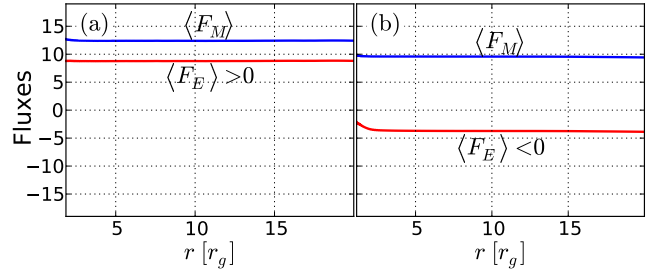}
  \end{center}
  \caption{Time-averaged fluxes of rest mass energy $\avg{F_M}$ and
    total energy $\avg{F_E}$ as a function of radius for models A0.5
    (panel a) and A0.99fc (panel b).  A positive flux means that the
    corresponding mass/energy flux is into the BH.  For model A0.5
    ($a=0.5$), both $F_M$ and $F_E$ are positive, i.e., inward.  The
    difference $(F_M-F_E)$ is the energy flux returned to
    infinity. The energy outflow efficiency
    (eq.~\ref{eq:eta2}) for this model is $\avg\eta\approx 30\%$,
    i.e., $30$\% of the accreted rest
    mass energy is returned as outflow energy.  For model A0.99fc,
    which has a rapidly rotating black hole ($a=0.99$), $\avg{F_E}<0$,
    i.e., net energy flux is out of the BH.  Correspondingly,
    $\avg{\eta} \approx 140\%$. Greater than $100$\%
    efficiency is possible only by extracting net energy from the
    spinning BH.  The gravitational mass of this BH decreases with
    time, though its irreducible mass increases.  The effects of
    density and internal energy floors have been subtracted from the
    calculated fluxes.  The subtraction is imperfect near the BH
    horizon, causing the slight non-constancy of fluxes at small $r$.}
 \label{fig:fluxvsr}
\end{figure}
Figure~\ref{fig:fluxvsr} shows plots of $\avg{F_M(r)}$ and
$\avg{F_E(r)}$ vs $r$ for the two simulations, A0.5 and
A0.99fc.  The fluxes have been averaged
over time intervals $(10300-13095)r_g/c$ and $(7000-30500)r_g/c$,
respectively, to reduce the effect of fluctuations due to flux
eruptions.  The time intervals have been chosen to represent
quasi-steady magnetically-arrested accretion.
The calculated fluxes are very nearly constant out to $r=20M$, indicating
that both simulations have achieved steady state in their inner
regions.  Consider first the results for model A0.5 with $a=0.5$
(Fig.~\ref{fig:fluxvsr}a). We find $\avg{F_M} \approx 12$ units
and $\avg{F_E}\approx 9$ units. 
The difference between these two fluxes
represents the energy returned by the accretion flow to the external
universe.
Our simulations are
non-radiative, with no energy lost via radiation.  Hence the energy
outflow is entirely in the form of jets and winds.

We define energy outflow efficiency $\eta$
as the
energy return rate to infinity
divided by the time-average rest mass accretion rate: 
\begin{equation}
\eta \equiv \frac{F_M-F_E}{\avg{F_M}} \times 100\%.
\label{eq:eta2}
\end{equation}
For model A0.5, the efficiency we obtain, $\avg{\eta}\approx 30\%$, 
is much
larger than the maximum efficiencies seen in earlier
simulations for this spin.
The key difference is that, in our simulation, we 
maximized the magnetic flux around the BH. This enables the system to
produce a substantially more {efficient} outflow.

In the more extreme model A0.99f with $a=0.99$
(Fig.~\ref{fig:fluxvsr}b), we find
$\avg{F_M}\approx 10$ units and
$\avg{F_E}\approx -4$ units.  The net energy flux
in this simulation is {\it out of} the BH, not {\it into} the BH, i.e.,
the outward energy flux via the Penrose/BZ mechanism
overwhelms the entire mass energy flux
flowing into the BH. Correspondingly, the efficiency
is greater than $100\%$: $\avg{\eta}\approx 140\%$.  Since
the system steadily transports net total energy out to infinity, the
gravitational mass of the BH decreases with time.  Where does the energy come
from? Not from the irreducible mass of the BH, which
cannot decrease in classical GR.  The energy comes from the
free energy associated with
the spin of the BH. The
BZ effect, which has efficiency $\etabh\approx135\%$ 
(eq.~\ref{eq:eta} with $\avg{\phibh^2}\approx47^2$ from
Fig.~\ref{fig:qtyvst}f and
$\kappa=0.044$), accounts for most
of the extracted energy.

Since greater than $100$\% efficiency has been a long-sought goal, we
ran model A0.99fc for an unusually long time ($t>30000r_g/c$).  There
is no indication that the large efficiency is a temporary fluctuation
(see Fig.~\ref{fig:qtyvst}g).  As a further check, we
calculated efficiencies for each of the runs, A0.99,
A0.99f, A0.99fh (Table \ref{tab1}), 
to estimate the uncertainty in $\eta$.  We
conclude that $\avg{\eta} \approx 140\pm15\%$ and that an
outflow efficiency $\gtrsim100$\% is achievable with a fairly realistic accretion scenario.
We note, however, that by changing the initial setup, e.g., the geometry of
the initial torus and the topology of the magnetic field, it might
be possible to obtain even larger values of $\avg{\eta}$.  This is an area for future
investigation.

Our outflows 
are in the form of twin
collimated relativistic jets along the poles and less-collimated
sub-relativistic winds \citep{lov76,bp82}. The former are mostly confined to streamlines
that connect to the BH, while the latter emerge mostly from the inner
regions of the accretion flow.  The bulk of the outflow power is in the
relativistic component.
The energy outflow efficiency shows considerable fluctuations
with time (Fig.~\ref{fig:qtyvst}g), reaching values as
large as $\eta\gtrsim 200$\% for prolonged periods of time, with a
long-term average value, $\avg\eta=140\pm15\%$.  
This may explain sources with very {efficient} jets
\citep{mcnamara_agnjetefficiency_2010,fernandes_agnjetefficiency_2010,punsly2011}.
The quasi-periodic nature of the fluctuations in $\eta$
suggests
magnetically-arrested accretion as a possible mechanism to produce
low-frequency QPOs in accreting stellar-mass BHs \citep{rem06} and
variability in AGN \citep{ghi_blazars_2010} and GRB outflows
\citep{progazhang2006}.  Additional
studies are necessary to ensure the convergence of variability
properties with numerical resolution.

We conclude that rapidly spinning BHs embedded in
magnetically-arrested accretion flows can produce efficient
outflows with $\avg\eta\gtrsim100\%$.  Such flows could be relevant for
understanding astrophysical systems with extremely {efficient} jets.
The fiducial model A0.99fc presented here, which is designed to
mimic magnetically arrested systems in nature, has a net energy flux
\emph{away} from the BH and  demonstrates
that net extraction of energy out of an accreting BH is viable via the Penrose/BZ effect.

\section*{Acknowledgements}
We thank S.\ Balbus, K.\ Beckwith, A.\ Benson, V.\ Beskin, I.\ Contopoulos, L.\ Foschini,
D.\ Giannios, J.\ Goodman, I.\ Igumenshchev, B.\ Metzger, R.\ Penna, R.\ Rafikov,
A.\ Spitkovsky, J.\ Stone, F.\ Tom\-besi, D.\ Uzdensky for discussions. We thank the
anonymous referee for useful suggestions. AT was supported by a Princeton
Center for Theoretical Science Fellowship. AT and RN were supported in part by
NSF grant AST-1041590 and NASA grant NNX11AE16G.  We acknowledge
support by the NSF through TeraGrid resources
provided by NICS Kraken, where simulations were carried out, 
NICS Nautilus, where data were analyzed, and NCSA MSS, where data
were backed up, under grant numbers
TG-AST100040 (AT), TG-AST080026N (RN) and TG-AST080025N (JCM).

{\small

\input{ms.bbl}
\label{lastpage}
}





\end{document}

%% file: macros.tex
\newcommand\araa{\rmfamily{ARA\&A}}%
\newcommand\apj{\rmfamily{ApJ}}%
\newcommand\apjl{\rmfamily{ApJ}}%
\newcommand\apss{\rmfamily{Ap\&SS}}%
\newcommand\mnras{\rmfamily{MNRAS}}%
\newcommand\na{\rmfamily{New Ast.}}%
\newcommand\pasj{\rmfamily{PASJ}}%
\newcommand\nat{\rmfamily{Nature}}%